\documentclass[aps,prl,twocolumn,showpacs]{revtex4}

\usepackage{graphicx} 
\usepackage{amsmath}
\usepackage{stmaryrd}
\usepackage{MnSymbol}
\usepackage{setspace}
\usepackage{color}

\begin{document}


\title{Spatial Coherence of Light in Collective Spontaneous Emission}
\author{D. C. Gold, P. Huft, C. Young, A. Safari, T. G. Walker, M. Saffman, and D. D. Yavuz}
\affiliation{Department of Physics, 1150 University Avenue,
University of Wisconsin at Madison, Madison, WI, 53706}
\date{\today}
\begin{abstract}

When a quantum system is put into an excited state, it will decay back to the ground state through a process termed spontaneous emission. It is generally assumed that spontaneous emission between different individual emitters would not be coherent with each other; to produce coherent light one would need population inversion and stimulated emission. In this work, we show that an optically-thin ensemble of 11,000 radiating atoms spontaneously organize to produce spatially coherent light. The reason for this coherence is collective-coupling of the individual emitters via Dicke superradiance and subradiance (as opposed to amplification through stimulated emission). 

\end{abstract} 
\maketitle

Spontaneous emission occurs due to the coupling of a quantum system (for example a neutral atom) to a continuum (infinite number) of radiation modes. It was predicted by Dicke 70 years ago that the usual process of spontaneous emission could be importantly modified when there is an ensemble of emitters \cite{dicke}. Specifically, the emission (decay) rate can be enhanced or reduced compared to the natural rate of a single isolated atom: effects commonly referred to as superradiance and subradiance \cite{haroche,scully1,scully2}. This enhancement and reduction of the decay rates can be understood classically as the constructive and destructive interference between the radiation originating from individual emitters. The physical picture is that when the emitters radiate in-phase (constructive interference), the decay rate is enhanced (superradiance); while out-of-phase emissions (destructive interference) result in reduced decay rates (subradiance). Whether there is superradiance or subradiance, spatial coherence is established between the atoms which is then mapped to their emitted light (i.e, the individual emitters are no-longer uncorrelated but instead have a defined phase relationship). This coherence is essential to collective decay: It is no coincidence that Dicke’s original paper is titled “{\it Coherence} in Spontaneous Radiation Processes”. It is this spatial coherence of the light that our experiment demonstrates. 

Superradiance is relatively easy to observe in dense ensembles that are in the Dicke limit, where many atoms are within a cubic wavelength of volume. In this regime, all that is required for superradiance to occur is to excite the atoms to a higher-energy level. For this case, due to the symmetries in the system, the ensemble primarily decays through in-phase superpositions. The first experimental observations of superradiance date to the early 1970s \cite{feld,manassah}. In contrast, the first studies of subradiance, especially in large ensembles, were not performed until much more recently \cite{kaiser1,kaiser2,browaeys}. These effects have been experimentally observed in a large number of physical systems including neutral atoms, ions, molecules, nitrogen vacancy centers in diamond, and superconducting Josephson junctions \cite{an,gauthier,kuga,thompson,molecules,ions,atoms,nanofiber,metamaterial,bec,havey,diamond1,diamond2,superconducting}. These experimental studies have been complemented by a large body of theoretical work \cite{eberly,scully3,adams,zoller,jenkins,ritsch,zanthier,agarwal}. We also note that, over the last decade, collective decay has found new applications in quantum computing and quantum information processing. For example, subradiance can be utilized to decrease the decoherence rate in a quantum system and therefore increase the lifetimes of stored quantum information \cite{kimble}. Using collective-coupling, two-dimensional arrays can be used for highly-directional mapping of quantum information between atoms and light \cite{yelin,bloch}.

In our experiment, we study spontaneous emission from a laser-cooled ensemble of rubidium ($^{87}$Rb) atoms. We observe subradiance in a regime that has not been studied before: a dilute (very small number of atoms per cubic wavelength), optically-thin cloud (thickness more than two orders of magnitude lower than previous experiments from other groups \cite{kaiser1,kaiser2,browaeys}, and an order of magnitude lower than our earlier work \cite{dipto}), in the strong-excitation regime. We show that in this regime, the subradiant time scales are not determined by the optical depth, but rather by the figure-of-merit for coherent emission. By coupling the radiated light to a misaligned Michelson interferometer, we show that for a sufficient number of atoms, the emitted light becomes transversely spatially coherent with a coherence length comparable to the size of the minor axis of the cloud. This coherence is  established in an optically-thin medium and is destroyed when the number of atoms in the ensemble is reduced. Furthermore, because the light coherence relies on appropriately phased or anti-phased superpositions in the atomic ensemble, it is quite sensitive to motional dephasing. Even a temperature increase of  $70$~ $\mu$K of the radiating atoms (corresponding to a motional dephasing of $\approx \lambda /200$ within the natural decay lifetime of $26.2$~ns) is enough to destroy the spatial coherence of the emitted light. 

Our results point to a new way of generating coherent light with characteristics distinctly different from a laser. Spatial coherence is established due to collective coupling of the atoms to radiated light and is produced in an optically-thin medium without any feedback (i.e., there is no external cavity). Furthermore, the light coherence in our experiment does not rely on amplification through stimulated emission in an inverted medium. The excited state fraction in our experiments is about 0.3 (i.e., there is no population inversion). 

\begin{figure}[tbh]
\vspace{0cm}
\begin{center}
\includegraphics[width=9cm]{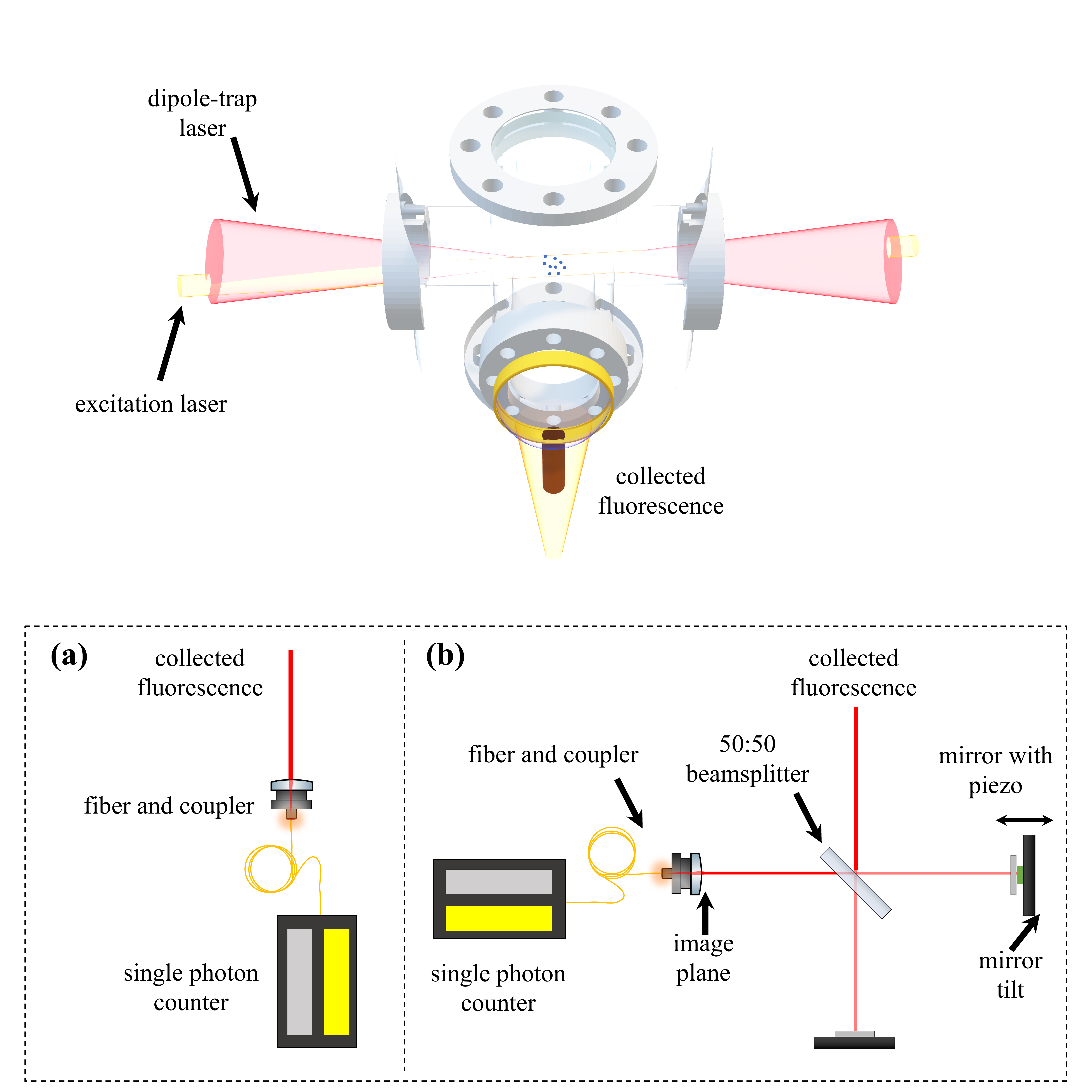}
\vspace{-0.2cm} 
\begin{singlespace}
\caption{\label{scheme} \small  Simplified experimental schematic. An ultrahigh vacuum chamber houses a magneto-optical trap (MOT) of laser-cooled ultracold rubidium ($^{87}$Rb) atoms. We overlap an intense far-off resonant laser at a wavelength of $1.055$~$\mu$m, trapping the atoms near the focus of the laser (an optical dipole trap). With the atoms confined to the dipole-trap and optically pumped to the $F=2$ ground level, we turn-off the dipole trapping laser and turn on a short intense laser pulse that couples the atoms strongly to the $F'=3$ excited level. After the excitation laser is switched-off, we collect the fluorescence from the atoms using a large lens and perform two types of studies: (a) we detect the emitted light with a single-photon counter and analyze decay as a function of time, and (b) we use a Michelson interferometer to study the spatial coherence properties of the emitted light. 
}
\end{singlespace}
\end{center}
\vspace{-0.cm}
\end{figure}

Figure~1 shows a simplified schematic of our experiment \cite{dipto}. The experiment starts with a magneto-optical trap (MOT) of laser-cooled ultracold rubidium ($^{87}$Rb) atoms, which is loaded from a background vapor inside an ultrahigh vacuum chamber. Laser cooling is implemented on the $F=2 \rightarrow F'=3$ cycling transition of the D2 line of $^{87}$Rb, near a wavelength of 780 nm. We overlap an intense far-off resonant laser at a wavelength of $1.055$~$\mu$m, trapping the atoms near the focus of the laser (an optical dipole trap). Further details of our experimental system can be found in Appendix~A. We typically trap 11,000 atoms at an atomic temperature of $40$~$\mu$K. Due to the profile of the trapping beam, the trapped atomic cloud is highly asymmetric with a radii of $6.3$~$\mu$m~$\times~6.3$~$\mu$m~$\times~360$~$\mu$m in the three spatial dimensions (all quoted numbers are the $1/e$ density radius).  With the atoms confined to the dipole-trap and optically pumped to the $F=2$ ground level, we turn-off the dipole trapping laser and turn on a single short intense laser pulse for 200 ns that couples the atoms strongly to the $F'=3$ excited level. This laser, termed the excitation laser, is spatially larger than the size of the ensemble and has a saturation parameter of $s/s_0 \approx 2$ (a discussion of the calibration of the saturation parameter can be found in Appendix~D). This saturation parameter corresponds to an excited state fraction of about 0.3 (i.e., each atom has a 30\% probability of being in the excited state). The excitation laser is linearly polarized in a direction orthogonal to the laser propagation direction as well as to the direction of fluorescence detection. We then abruptly turn-off the excitation beam and observe the fluorescence emitted from the atoms. The abrupt switching of the excitation laser is achieved using a fast acousto-optic modulator with a 90\%-10\% turn-off time of 8 ns, and allows us to observe purely spontaneous emission by the atomic ensemble.

With the excitation laser switched-off, the atoms spontaneously decay to their ground level. The lifetime of the excited level for an isolated atom is $\tau_a=26.2$~ns, and this quantity sets the relevant time-scale for our experiment. We collect spontaneous emission from the ensemble with a lens of numerical aperture NA$=0.19$, in a direction orthogonal to the propagation direction of the excitation laser. The emission in the orthogonal direction does not carry any correlations imprinted due to the phase-fronts of the laser. This is in contrast to temporal coherence studies that look along the forward direction, typically referred to as Free Induction Decay \cite{induction}. We perform two types of experimental studies with the collected light. (i) In the first type, we measure the collective decay of the ensemble as a function of time and study subradiance [inset (a)]. For this purpose, we detect the collected light with a single-photon counting module, and average over many experimental cycles, which gives us a time-resolved trace of the spontaneous decay. (ii) In the second type of study, we couple the collected light to a misaligned Michelson interferometer [inset (b)]. Here, we split the beam into two arms using a 50/50 beam-splitter. Each arm reflects off a mirror and then the two arms are recombined with one slightly offset from the other and detected. This set-up enables us to measure the coherence of the emitted light across its spatial profile since the collection lens produces an image of the radiated light from the ensemble near the detector of the interferometer. 

\begin{figure}[tbh]
\vspace{-0cm}
\begin{center}
\includegraphics[width=9cm]{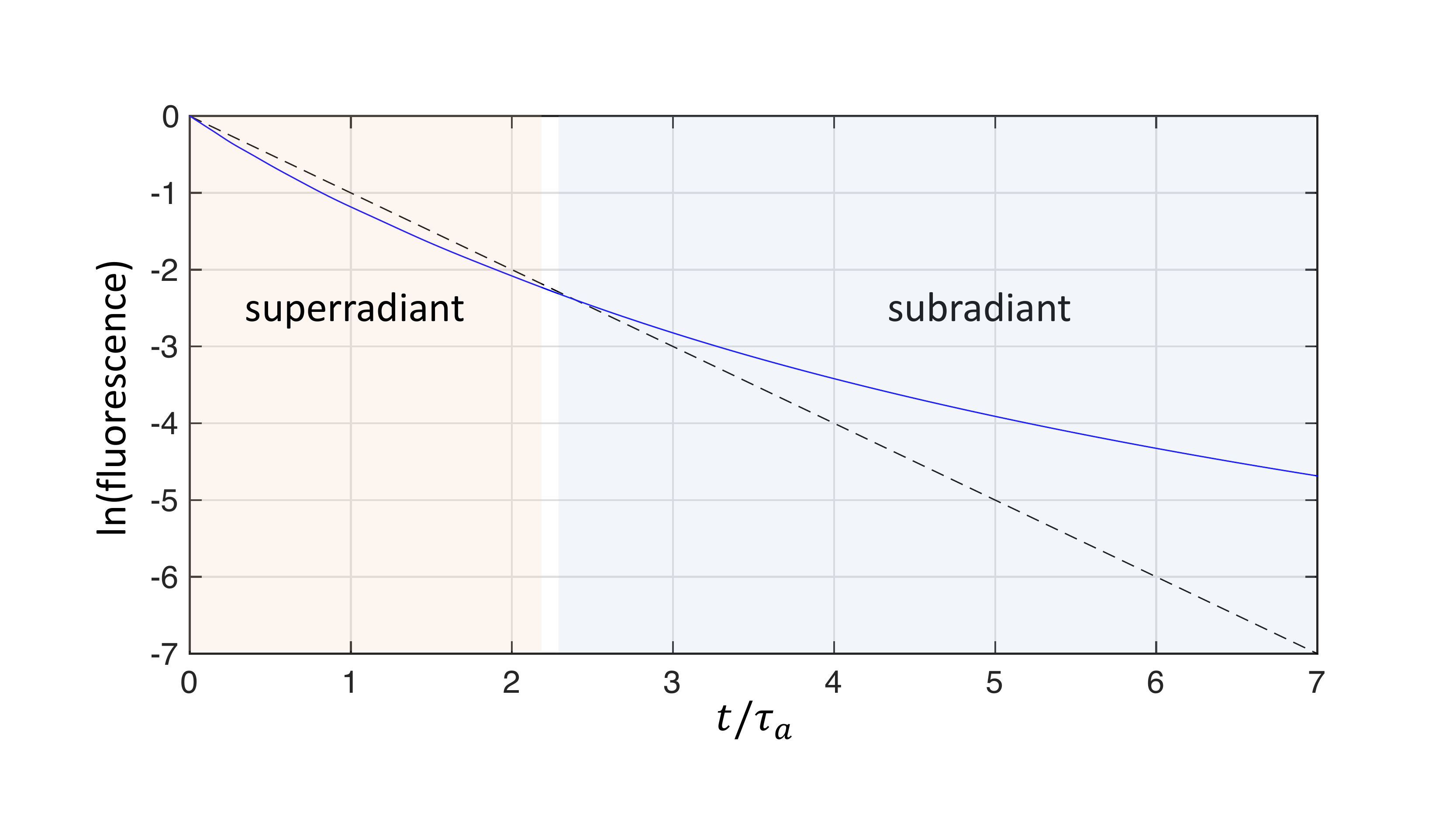}
\vspace{-1cm} 
\begin{singlespace}
\caption{\label{scheme} \small  A sample numerical simulation for parameters that are similar to what we have in our experiment. Here, we use the Dicke decay ladder model modified by the eigenvalue distribution of each subspace, as discussed in Refs. \cite{dipto,ben}. The solid blue line shows the fluorescence from the ensemble in logarithmic scale while the dashed black line is the decay of an isolated atom, $\exp{(-t/\tau_a)}$. There is initial superradiance followed by subradiant decay. While the initial superradiance is difficult to observe in dilute samples, the subradiance is more easily observable, since it produces large deviations from independent decay at later times. 
}
\end{singlespace}
\end{center}
\vspace{-0.cm}
\end{figure}

The fact that these effects can be present in a dilute, disordered, and large sample is not obvious, and here we provide a qualitative explanation. Initially, right after the excitation laser beam is turned-off, light emitted along the direction of the detector (orthogonal to the laser propagation direction) comes from random locations (since the atomic ensemble is disordered) and does not display any phase coherence. However, some of the atoms are near the correct positions so that their emissions interfere constructively, producing superradiance (faster than the independent emission rate). These correlated superradiant modes (and uncorrelated independent emission modes) decay  relatively quickly, leaving only anti-phased superpositions in the sample, which in turn produce subradiance. For this process to happen, it is critical that the atoms are almost stationary during the emission process so that their phase relations are preserved: Ultracold temperatures are, therefore, essential. The result is spatially coherent light formed purely due to spontaneous emission: early in the decay this manifests predominantly through phased superpositions, while later stages are dominated by anti-phased superpositions. The amount of spatial coherence is determined by the proportion of atoms emitting through correlated channels compared to those emitting through independent, uncorrelated modes. Figure~2 shows a sample numerical simulation with parameters that are similar to what we have in our experiment. Here, we use the Dicke decay ladder model modified by the eigenvalue distribution of each subspace, as discussed in Refs. \cite{dipto,ben}. The solid blue line shows the fluorescence from the ensemble on a logarithmic scale while the dashed black line is the decay of an isolated atom, $\exp{ (-t/ \tau_a)}$. Two distinct regimes of the decay, superradiant and subradiant, are clearly evident. For the conditions of our experiment (a dilute cloud with a very-low optical depth) the initial superradiance is difficult to observe since the superradiant time evolution is not long enough to produce large deviations from independent decay. However, the subradiance is more easily observable, since it persists for longer time scales. 

\begin{figure}[tbh]
\vspace{-0cm}
\begin{center}
\includegraphics[width=9cm]{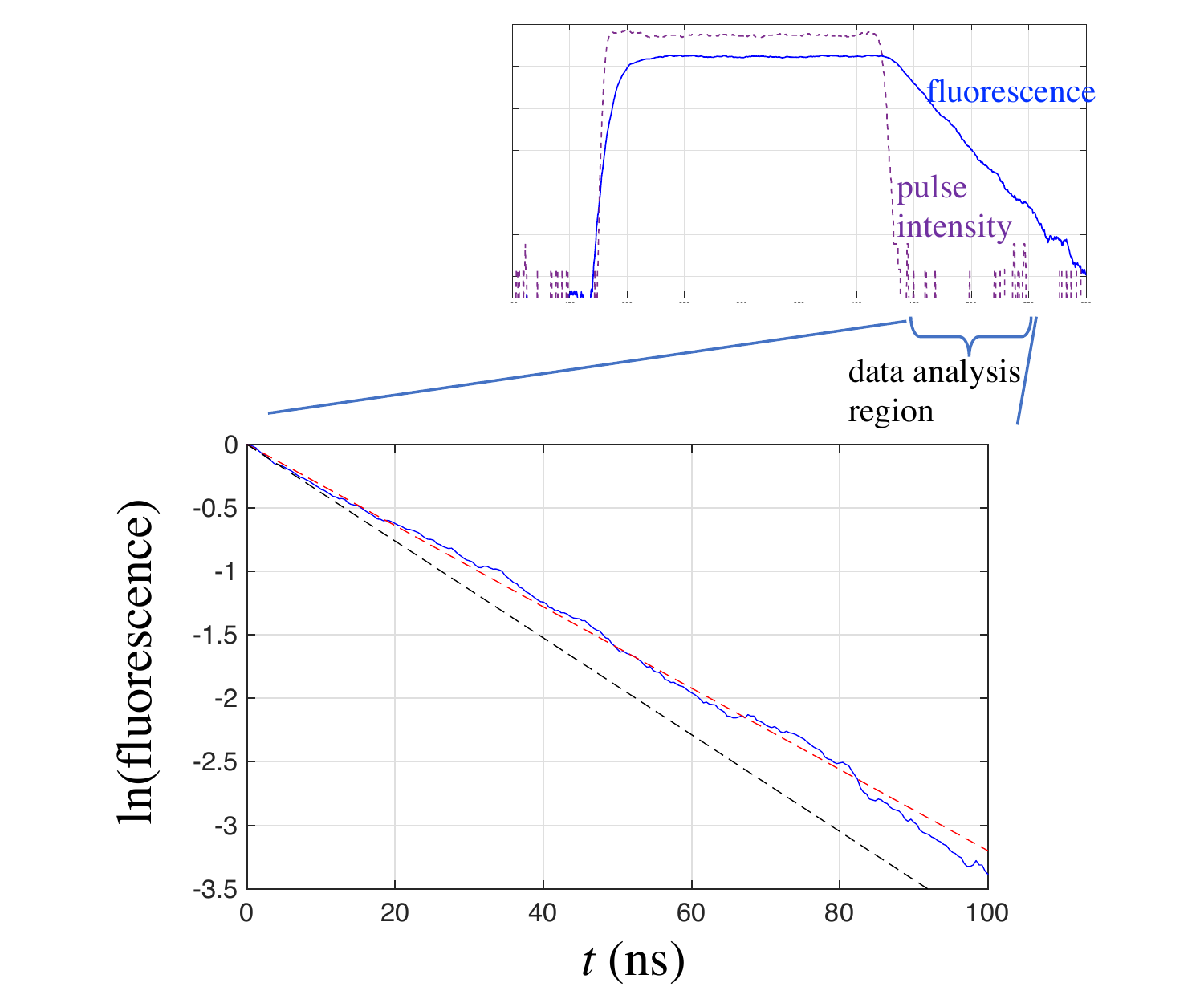}
\vspace{-1cm} 
\begin{singlespace}
\caption{\label{scheme} \small  An example decay curve (solid blue line) of the fluorescence where we observe substantial subradiance. For this example, which is obtained for an atomic density of $10^9$/cm$^3$, the $1/e$ decay constant is increased by about 20\%, i.e. $\tau = \tau_a$. The dashed black line is the expected decay of an isolated individual atom for comparison, while the dashed red line is the linear fit to the observed data. The inset shows a sample fluorescence trace (solid blue curve) as well as the intensity of the excitation pulse as detected on a fast photo-detector (dashed purple curve) on a longer time span. All the data that we will present below, including the coherence measurements on the Michelson interferometer, are done in the data analysis region marked in the inset; i.e., after the excitation pulse is switched off.
}
\end{singlespace}
\end{center}
\vspace{-0.cm}
\end{figure}

We proceed with a detailed description of our experimental results. For each experimental cycle, the number of detected photons on the counter is of order unity. For our subradiance studies, we re-prepare the ensemble and average over $\sim 10^5$ experimental cycles. Each experimental cycle (MOT loading, dipole-trap loading, excitation, fluorescence detection) takes about $1$~second. As a result, each fluorescence decay trace requires a data acquisition time of about a day. Figure~3 shows an example decay curve of the fluorescence in which we observe substantial subradiance. For this decay curve, the $1/e$ time constant is increased by about 20\% to $\tau = 1.2 \tau_a $. As we discuss below, this increase in the time constant cannot be due to incoherent processes such as radiation trapping due to the very low optical depth of the sample \cite{kaiser2}. The inset of Fig.~3 shows a sample fluorescence trace (solid blue curve) as well as the intensity of the excitation pulse for reference (dashed purple curve), both plotted on a logarithmic scale. All of the data that we present below, including the spatial coherence measurements on the Michelson interferometer, are recorded in the “data analysis region” marked in the inset (i.e., after the excitation pulse is switched off and the atoms are in free space). 

\begin{figure}[tbh]
\vspace{-0.5cm}
\begin{center}
\includegraphics[width=9cm]{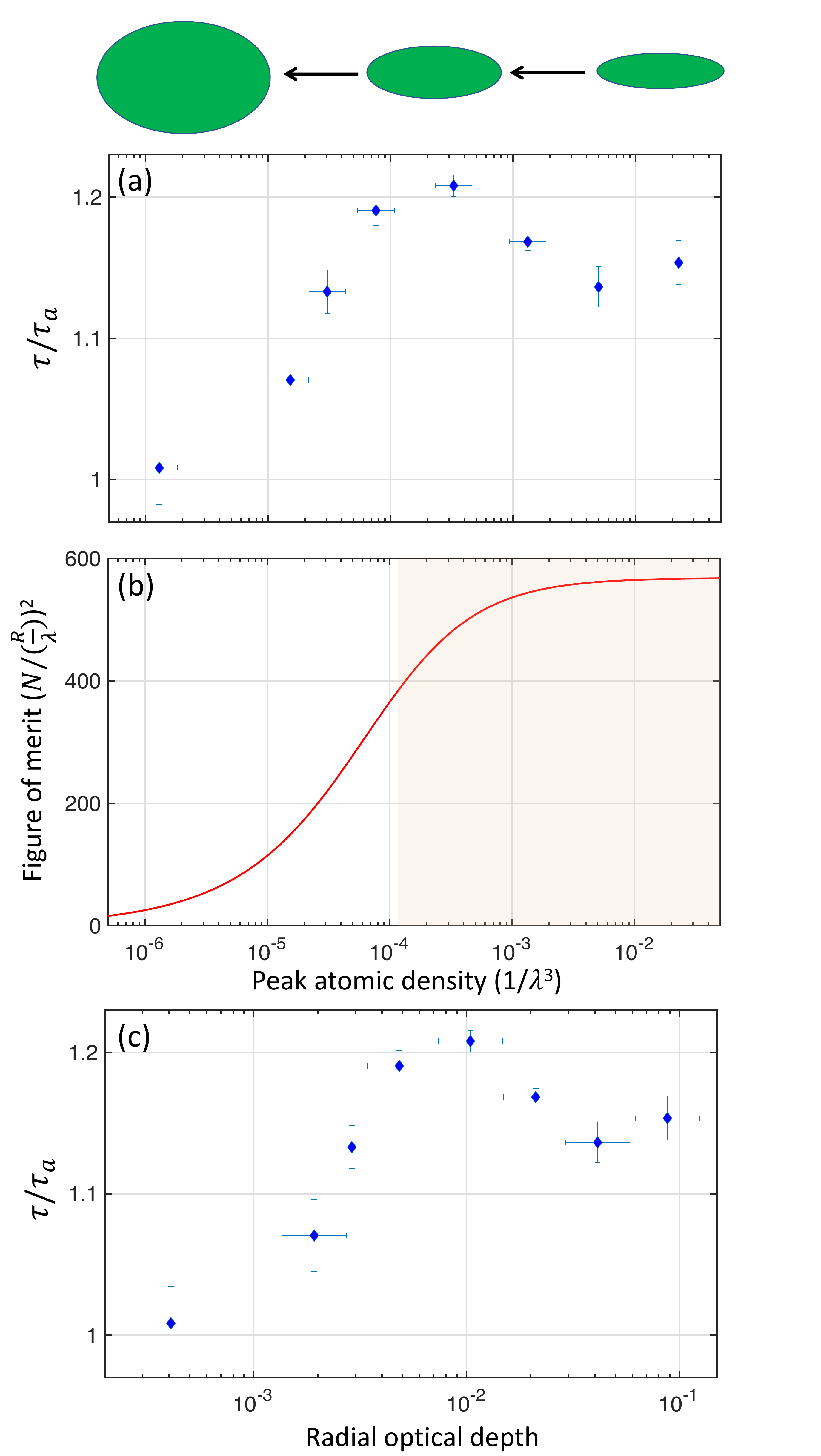}
\vspace{-0.5cm} 
\begin{singlespace}
\caption{\label{scheme} \small  (a) The observed subradiance as a function of density of the sample.  At low densities, the decay time approaches the independent decay time, $\tau \rightarrow \tau_a$. As the density is increased, the observed subradiance increases and then reaches a plateau, followed by a slight drop near the highest density. (c) The same data points plotted as a function of optical depth. The general behavior of this data (subradiance increasing and then reaching a plateau) is expected from the figure-of-merit that would describe coherent emission from the ensemble, $FoM \equiv \left( \frac{N} {R/ \lambda} \right)^2$. This figure-of-merit is plotted in (b). }
\end{singlespace}
\end{center}
\vspace{-0.5cm}
\end{figure}

Figure~4(a) shows the observed decay time as a function of the peak density of the ensemble. We vary the density by allowing the atomic cloud to freely expand for a given amount of time after the dipole-trap is shut-off and before the excitation pulse is applied. The data points are obtained for cloud expansion times of $0.1$, $0.25$, $0.5$, $1$, $2$, $3$, $4$, and $10$ ms, respectively. Here, for each expansion time, we fit a curve to the data (for a duration of $3.5 \tau_a$ after the excitation beam is turned-off) and find the 1/e decay-time of the fit, which we refer to as $\tau$. The vertical error bar on each data point is a measure of the uncertainty in the fit. The horizontal error bars are primarily due to the uncertainty in the measurement of the number of atoms in the sample. At low densities, the decay time approaches the independent decay time, $\tau \rightarrow \tau_a$, as expected. As the density is increased, the observed subradiance increases and then reaches a plateau, followed by a slight drop near the highest density. The drop is likely due to a currently unidentified process which dephases the superpositions across the ensemble. Figure~4(c) displays exactly the same data points but plotted as a function of the radial optical depth (the optical depth of the cloud along the direction of the detector). The optical depth of the sample where we observe the largest subradiance (a depth of $\approx 10^{-2}$) is more than two orders of magnitude less than in the systems where subradiance had previously been studied \cite{kaiser1,kaiser2,browaeys}.

The general behavior of the data of Fig. 4 (subradiance increasing with density and then reaching a plateau) is expected from the figure-of-merit that would describe coherent emission from the ensemble, $FoM \equiv \left( \frac{N} {R/ \lambda} \right)^2$. Here $N$ is the number of atoms in the sample, $R$ is the radial size of the cloud and $\lambda =780$~nm is the wavelength of the emitted light. This figure-of-merit is plotted in Fig.~4(b). The plateau in Fig.~4(b) is due to the highly asymmetric initial shape of the cloud that results from the shape of the dipole-trap beam at the focus. This is shown in the cartoon at the top of Fig.~4. Given this large aspect ratio, the initial expansion of the cloud (which is uniform in all three directions) does not change the overall size (the length of the major axis) appreciably. A qualitative derivation of the figure-of-merit for coherent emission, $\left( \frac{N} {R/ \lambda} \right)^2$, is given in Appendix~B. In contrast, for incoherent emission from the ensemble, the relevant figure-of-merit is $ \frac{N} {\left( R/ \lambda \right)^2} \sim n \sigma R$, i.~e., the optical depth of the sample ($n$ is the density and $\sigma$ is the on-resonant cross-section for light absorption). In our experiment, the portion of the data in the shaded region of Fig. 4(b) is particularly important. In this region, the atomic density varies by more than two orders of magnitude and the optical depth changes by more than an order of magnitude, and yet the observed subradiance is largely unchanged. This is a clear indication that the process is dominated by neither the optical density, nor the optical depth, but instead by the figure-of-merit due to coherent emission.

We next proceed with a discussion of our spatial coherence measurements. The collection lens forms an image of the fluorescing ensemble of atoms at the Michelson interferometer. This image is then split into two arms using a 50/50 beam-splitter. Each arm reflects off a mirror and the two arms are recombined back on the same beam-splitter. We detect the recombined signal with a photon counter. Using this set-up, we essentially interfere two separate images of the ensemble at the detector. With the mirrors, these two images can be displaced from each other in the transverse plane, allowing a measurement of the transverse spatial coherence. Any interference between the images can be characterized by precisely changing one of the arm lengths (in the longitudinal direction). We control the length of one of the arms by using a high-voltage piezo-electric transducer attached to one of the mirrors. As in our previous measurements, for all the interference and spatial coherence measurements that we discuss below, we only collect photons after the excitation beam is switched-off (indicated region in Fig. 3). All the Michelson interferometer measurements that we discussed below are obtained with the cloud freely expanding for a duration of 0.2 ms before the excitation beam is applied. The size of the cloud for the Michelson measurements is $18$~$\mu$m~$\times~18$~$\mu$m~$\times~360$~$\mu$m (all dimensions are the $1/e$ density  radius).  

\begin{figure}[tbh]
\vspace{-0cm}
\begin{center}
\includegraphics[width=9.5cm]{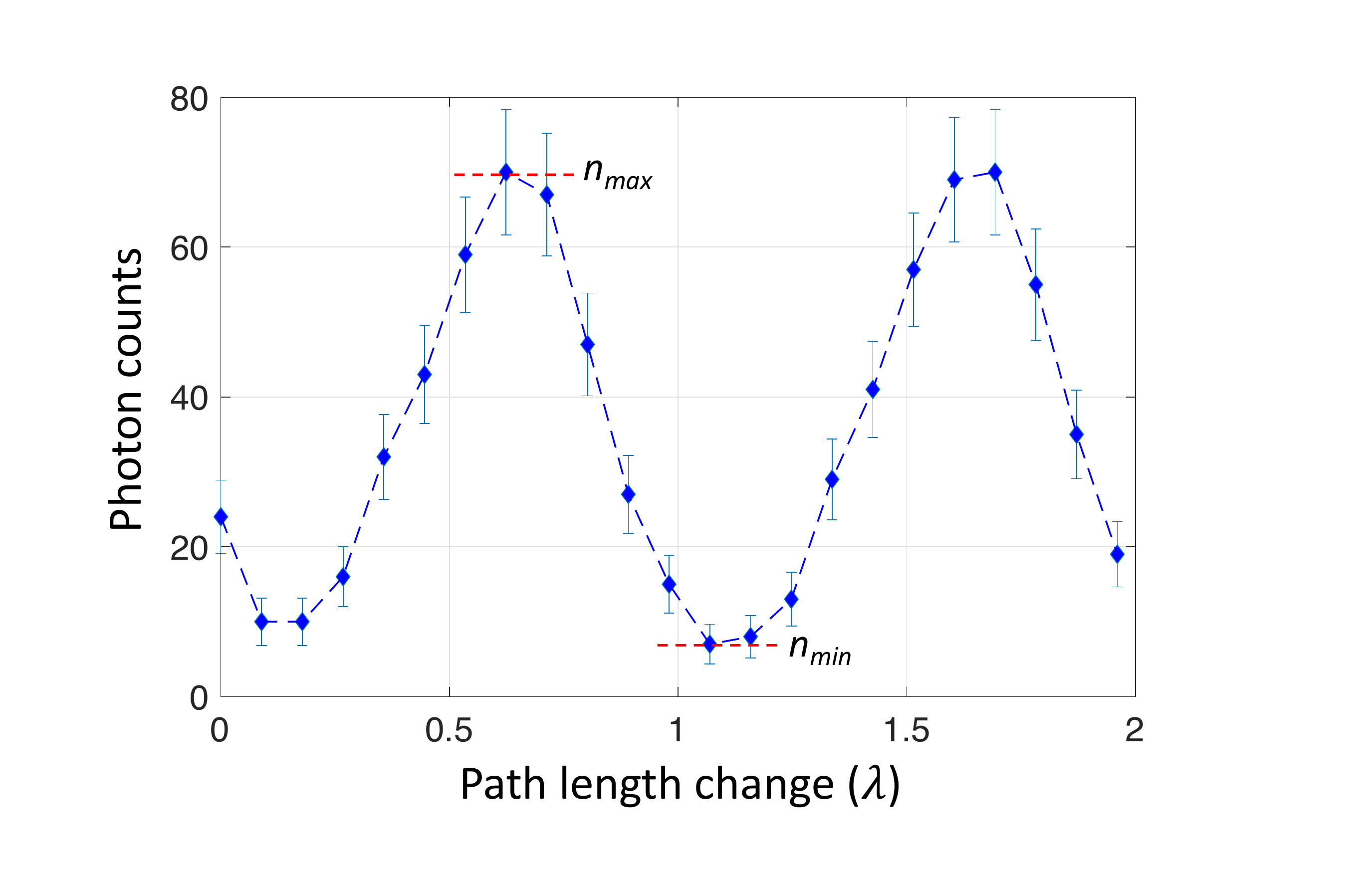}
\vspace{-1cm} 
\begin{singlespace}
\caption{\label{scheme} \small  A typical interference pattern that we observe for our optimal experimental conditions when the interferometer is well-aligned and the atomic density is high with $FoM \approx 550$. Here, we record the number of photons detected at the output of the Michelson interferometer as a function of change in the length of one of the arms. With a recorded interference pattern as shown here, we can use its contrast, defined by $(n_{max}-n_{min})/(n_{max}+n_{min})$, as a measure of the coherence between the two arms. }
\end{singlespace}
\end{center}
\vspace{-0.3cm}
\end{figure}

Figure 5 shows a typical interference pattern that we observe for our optimal experimental conditions. Here, we record the number of photons as a function of change in the length of one of the arms. Each data point is an average of 1000 experimental cycles. This particular trace is taken when the atomic density is high with $FoM \approx 550$ and the interferometer is well-aligned (i.e., images of the fluorescence coming from the two arms overlap within a few microns of each other). With a recorded interference pattern as shown in Fig. 5, we can use the contrast of the fringe pattern, defined by $ (n_{max}-n_{min})/(n_{max}+n_{min})$, as a measure of the spatial coherence between the two images. Here, the quantities $n_{max}$ and $n_{min}$ are the number of detected photons at the interference maxima and minima, respectively.  

\begin{figure}[tbh]
\vspace{0cm}
\begin{center}
\includegraphics[width=8cm]{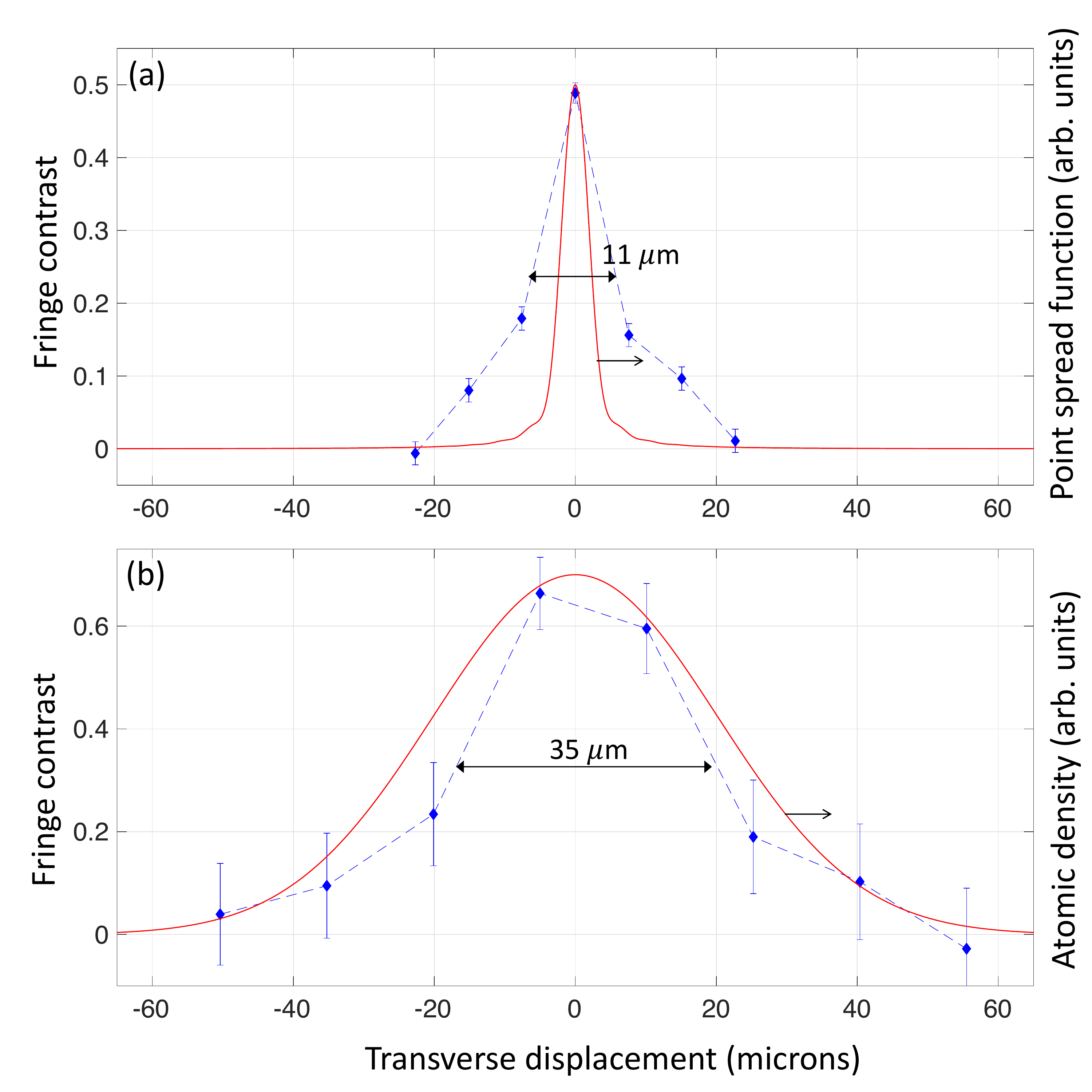}
\vspace{-0.2cm} 
\begin{singlespace}
\caption{\label{scheme} \small  The spatial coherence measurement (a) for the low-density ($FoM<1$)  and (b) high-density ($FoM \approx 550$) atomic cloud. Here we record the fringe contrast as a function of transverse displacement between the two arms of the interferometer. For the low-density case, we observe a full-width-at-half maximum (FWHM) of 11 microns. Theoretically calculated point spread function for our imaging system is also plotted in (a) for comparison (solid red line). For the high-density case of (b) the transverse coherence length increases to 35 microns. For comparison, calculated density profile of the radiating ensemble (based on the known initial size, atomic temperature, and expansion time), is also plotted (solid red curve). The observed spatial coherence length agrees well with the size of the atomic cloud; which shows that the spatial coherence of light largely extends to size of the minor axis. }
\end{singlespace}
\end{center}
\vspace{-0.3cm}
\end{figure}

Figure~6 shows the spatial coherence measurement for a low-density atomic cloud with $FoM<1$ [Fig. 6(a)] and a high-density cloud with $FoM \approx 550$ [Fig. 6(b)]. Here we record the fringe contrast as a function of transverse displacement between the two arms of the interferometer. For the low-density case, we observe a full-width-at-half maximum of 11 microns. Even with completely uncorrelated emission one would still expect a transverse spatial coherence length that is determined by the point spread function of the imaging system. This is because, each point source (individual atom) in the ensemble produces a spread in the image plane (as determined by the point spread function). The width of the point spread function is set by the wavelength of light and the numerical aperture of the collection lens. A theoretically calculated point spread function for our imaging system is also plotted in Fig. 6(a) for comparison (solid red line). The data points agree with the theoretical calculation reasonably-well; the discrepancy is likely due to imaging aberrations as well as imperfect alignment of the interferometer. For the high-density case of Fig. 6(b) the transverse coherence length increases to 35 microns. For comparison, the calculated density profile of the radiating ensemble (based on the initial size, atomic temperature, and expansion time), is also plotted (solid red curve). The observed spatial coherence length agrees well with the size of the atomic cloud; which shows that the spatial coherence of light becomes comparable to the size of the cloud along this axis.

\begin{figure}[tbh]
\vspace{-0cm}
\begin{center}
\includegraphics[width=9.5cm]{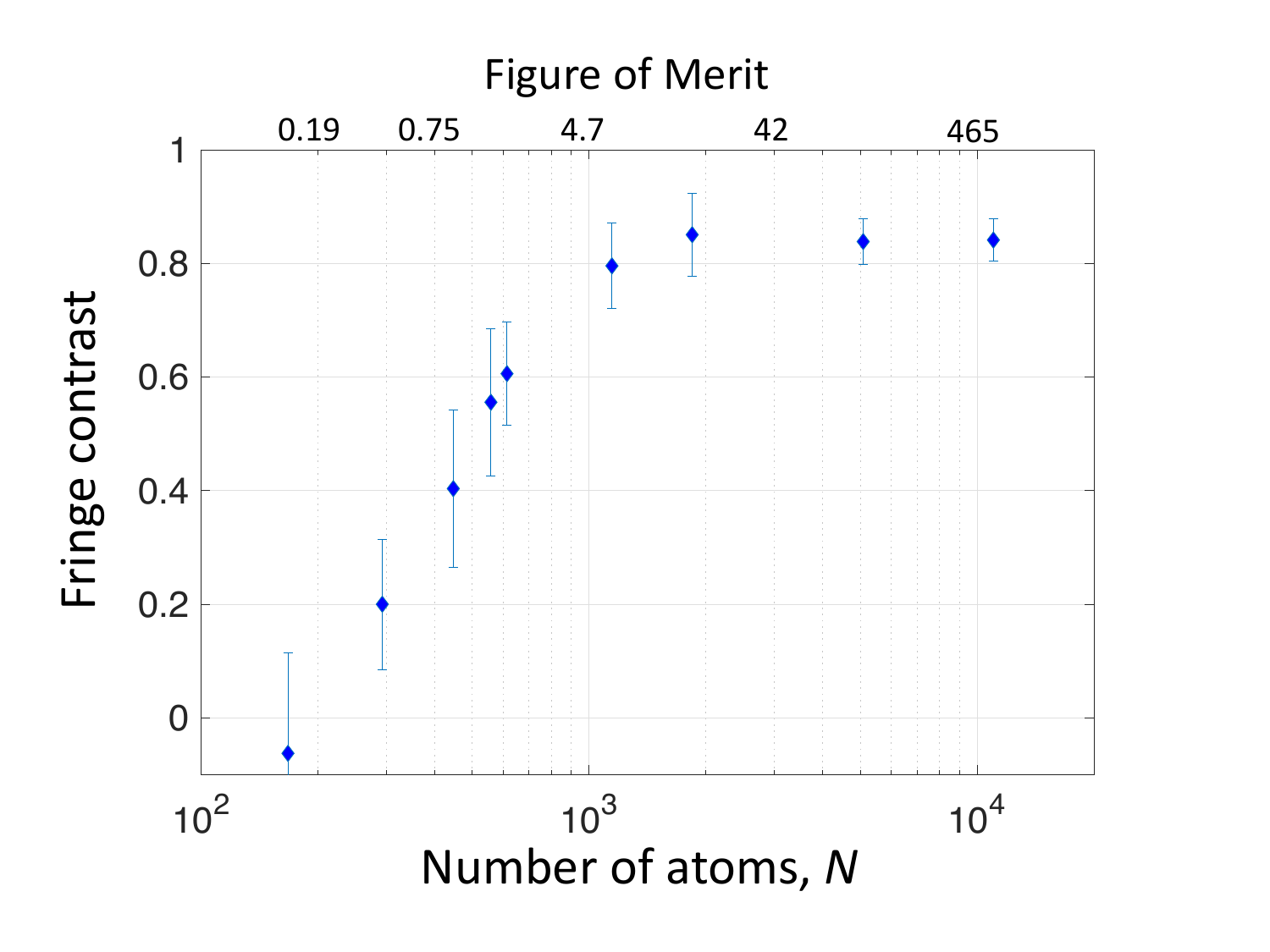}
\vspace{-1cm} 
\begin{singlespace}
\caption{\label{scheme} \small  The contrast of the interference fringes as a function of the number of atoms in the ensemble. Here, all the other parameters of the radiating ensemble, including its size ($18$~$\mu$m~$\times~18$~$\mu$m~$\times~360$~$\mu$m) and atomic temperature ($\approx 40$~$\mu$K) remain fixed. We vary the atom number, by changing the loading efficiency from the MOT to the dipole trap. As the number of atoms is reduced, the fringe contrast remains high up until about 1000 atoms, below which the contrast drops sharply. }
\end{singlespace}
\end{center}
\vspace{-0.3cm}
\end{figure}

Because there is such a large difference between the spatial coherence length for the high-density atomic cloud and the size of the point-spread-function, we can slightly misalign the two images in the interferometer (by about 10 microns) and measure the fringe contrast as we vary experimental parameters. Figure 7 shows the contrast of the interference fringes as a function of the number of atoms that are trapped in the dipole-trap. Here, all the other parameters of the radiating ensemble, including its size and atomic temperature remain fixed. We vary the atom number by changing the loading efficiency from the MOT to the dipole trap. As the number of atoms is reduced, the fringe contrast remains high down to about 1,000 atoms, below which point the contrast drops sharply. This clearly shows the collective nature of the observed coherence: One needs a sufficient number of atoms (and therefore sufficient collective coupling), for the emission to be spatially coherent.

\begin{figure}[tbh]
\vspace{-0cm}
\begin{center}
\includegraphics[width=9.5cm]{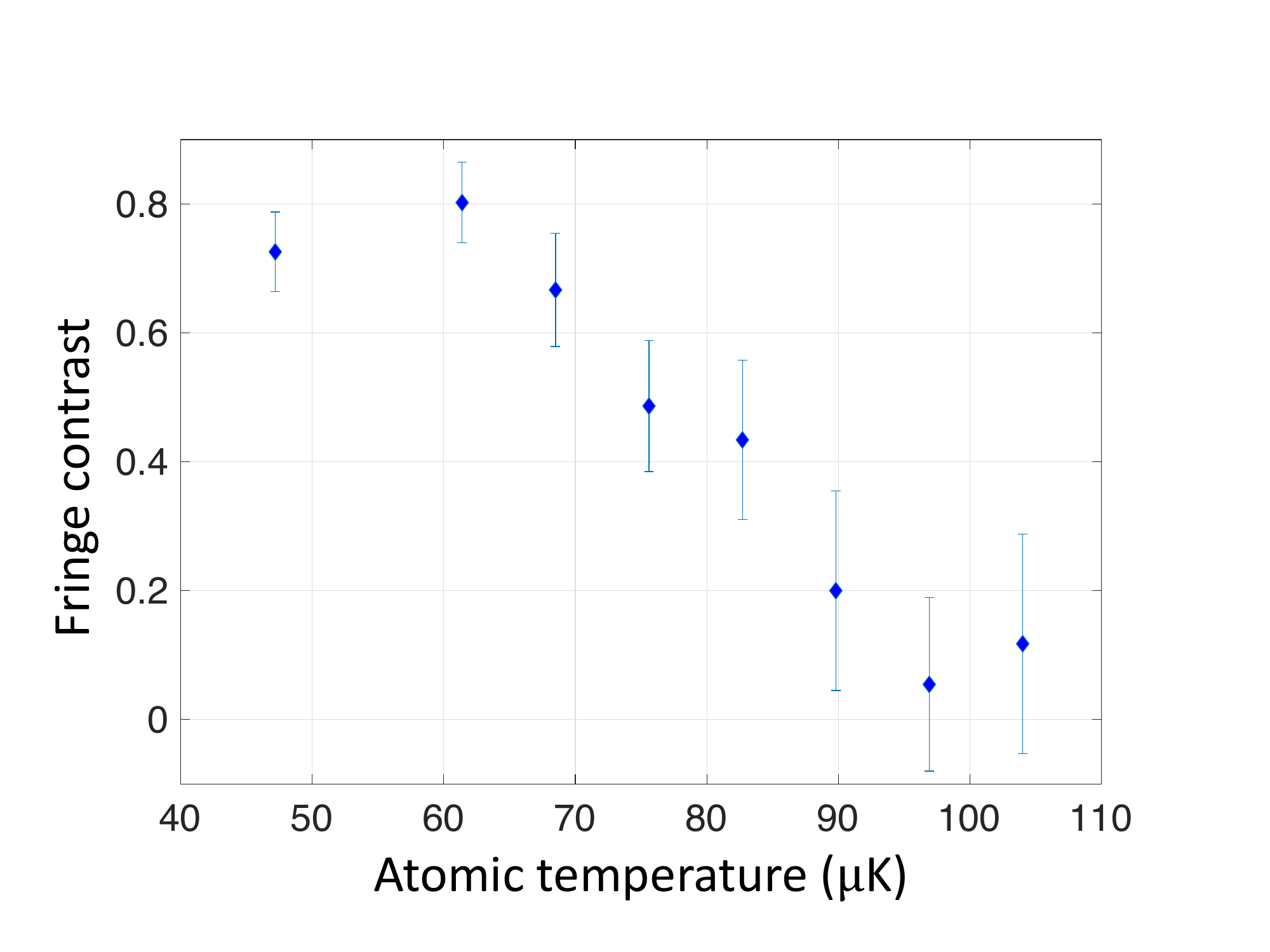}
\vspace{-1cm} 
\begin{singlespace}
\caption{\label{scheme} \small  The contrast of the interference fringes as a function of atomic temperature (with atom number fixed to 11,000 and the size of the cloud fixed to $18$~$\mu$m~$\times 18$~$\mu$m~$\times 360$~$\mu$m). We increase the atomic temperature by illuminating the atoms with an on-resonant laser pulse of varying duration. The fringe contrast vanishes, as the atomic temperature is increased from $45$~$\mu$K to $100$~$\mu$K.}
\end{singlespace}
\end{center}
\vspace{-0.3cm}
\end{figure}

In Figure 8, we keep the number of atoms and the size of the ensemble fixed (i.e., the figure-of-merit is fixed with $FoM \approx 550$) and vary the atomic temperature. We increase the atomic temperature by illuminating the atoms with an on-resonant laser pulse of varying duration while they are confined to the dipole trap. The fringe contrast vanishes as the atomic temperature is increased from 45~$\mu$K to 100~$\mu$K. This shows the sensitivity of the spatial coherence of the emitted light to motional dephasing. As mentioned above, the established coherence relies on maintaining phased and anti-phased superpositions in the atomic cloud during the spontaneous decay process. It is critical that these superpositions do not decohere during the relevant time-scale of the experiment, which is the decay time $\tau_a$. An atomic temperature of 100~$\mu$K corresponds to an average speed of $15$~cm/s for the atoms. This speed results in a motional dephasing of $\lambda /200$ during a decay time $\tau_a$, which, remarkably, is large enough to destroy the established coherence. 

The sensitivity to motional dephasing at the level of $\lambda /200$ can be qualitatively understood through the scaling of the collective rates with the number of atoms, $N$, in the ensemble. The eigenvalue distributions of the exchange Hamiltonian that describes collective decay in the large sample, dilute limit was derived in Ref. \cite{ben}. In a three dimensional geometry, the width of the eigenvalue distributions, which sets the superradiant and subradiant decay rates, scales as $N^{2/3}$. This essentially means that the vast majority of eigenvalues (and therefore the decay rates) are determined by the coherent interference of $N^{2/3}$ atoms. (Here, the basic idea is similar to the classical explanation of traditional Dicke superradiance. When there are $N$ emitters within a wavelength of light radiating in-phase, this results in a factor of $N$ enhancement in the total radiated power compared to incoherent emission (since the waves coherently add up), which then results in a factor of $N$ increase in the decay rate).  With analogy to a Fabry-Perot resonator, when $N^{2/3}$ emissions from individual atoms interfere, one would expect a sharpness in contrast versus motional dephasing of order $\lambda /N^{2/3}  \approx \lambda/494$, which is reasonably consistent with our experimental observation (Fig. 8). 

In conclusion, we have directly observed spatial coherence of the emitted light from the collective decay of an atomic ensemble. The observed coherence is over a time window of $3.5 \tau_a$ after the excitation beam is switched-off. We note that we currently do not have a good understanding of the precise spatio-temporal evolution of the coupled quantum states, and thus of the spatial structures of superradiant and subradiant modes that are contributing to the emission. In the regime of our experiment (a large sample with strong excitation), numerically calculating even a single superradiant or a subradiant mode (eigenvector) is an intractable problem due to the exponentially large size of the Hilbert space \cite{ben}. In the future, with increased signal to noise and using a detector array, it might be possible to image the spatial interference pattern of the Michelson interferometer at specific time points during the decay. Such time-resolved measurements could give information regarding the spatial profiles of the superradiant and subradiant modes, and thereby give insight into the structure of the Hilbert space. 

Another future direction would be to investigate the quantum statistical properties of the emitted radiation \cite{coherence}. Our preliminary studies indicate that photon statistics of our source can deviate substantially from Poissonian statistics that would be expected for independent (uncorrelated) atoms. Specifically, under certain conditions, we have preliminary experimental results that show photon-number squeezing (i.e., sub-Poissonian statistics). We also observe that the deviation from the Poisson distribution is correlated with the amount of subradiance that we observe. This is expected since the stronger the observed subradiance, the more correlated the system becomes. In future work, the squeezing in photon number could be characterized as the parameters of the atomic ensemble, such as the number of atoms, is varied. Furthermore, by beating the spontaneously emitted light with a local oscillator (a strong laser beam) in a homodyne interference set-up, it will be possible to detect quantum fluctuations in both the phase and the amplitude of the emitted light (i.e., in the quadrature amplitudes). 

Spontaneous emission is ubiquitous and fundamental, and is unavoidable when a quantum system is coupled to an environment with infinite degrees of freedom. Within the context of a quantum computer, spontaneous emission between the qubit levels will produce noise on the qubits. Our experiment shows that even in a large sample, such noise quickly transitions to being correlated across the ensemble. In current physical implementations of quantum computing \cite{kimble_review,NielsenChuang}, the spontaneous emission rate between the qubit levels can be very small (for, example, hyperfine qubits in neutral atoms are known to be very stable \cite{neutral_review1}). However, for scalability discussions, the small value of the decay rate is not relevant; rather the scaling of the correlated noise with the number of qubits ($N$) is the important quantity. Within this context, another future direction would be to explore the implications of our results in fault-tolerant quantum computation and the threshold theorem \cite{burkard,preskill,mucciolo1,mucciolo2}. 

More specifically, in our recent paper, we have discussed the effect of correlated noise due to collective spontaneous emission on fault-tolerance of quantum computing  \cite{ben}. The key idea is that collective spontaneous emission produces noise on each qubit with a strength that scales with the total number of qubits in the computer. As a result, the threshold theorem is violated because one cannot assume the error on each qubit during a gate time to be smaller than a certain constant threshold (which is independent of the total number of qubits in the computer). We believe that the temperature sensitivity of the observed spatial coherence (Fig. 8) provides a promising line of research along verifying the $N$ scaling of the noise due to collective spontaneous emission. Specifically, the temperature scans of the fringe contrast (similar to the measurement of Fig.~8) as the number of the atoms in the ensemble is varied, can reveal the expected $N^{2/3}$ dependence of the rates. With the $N$ scalings verified, avenues to eliminate correlated noise due to colllective emission can be explored.

\section{Acknowledgments}
We thank Dipto Das and Ben Lemberger for many helpful discussions. This work was supported by NSF Award 2016136 for the QLCI center Hybrid Quantum Architectures and Networks.

\section{Appendix A: Experimental Details}

{\it Magneto-optical trap (MOT)}: The experiment is performed inside an ultrahigh vacuum chamber which is kept at a base pressure of $\sim 10^{-9}$~torr. To form the $^{87}$Rb MOT, we use three counter-propagating beam pairs that are locked to the $F=2 \rightarrow F'=3$ cycling transition in the D2 line (with a transition wavelength of 780 nm). Two beam-pairs each have an optical power of about 40 mW and a beam radius of 3 cm. Due to space constraints, the third beam pair is not orthogonal to the other two, and so is smaller, with a beam radius of 5 mm, and an optical power of 5 mW. The MOT lasers are produced by a custom-built external-cavity diode laser (ECDL) whose output is amplified by a semiconductor tapered amplifier, before being split into beam pairs. Further details regarding our laser system can be found in our prior publications \cite{dipto}. The MOT lasers are overlapped with a hyperfine repumping beam, which is generated by a separate ECDL locked to the $F=2 \rightarrow F'=2$ transition with an optical power of about 1 mW.  

We load the atoms to the MOT from the background vapor in the chamber for about 400 ms. For the last 40 ms of loading, we detune the MOT lasers by about $8 \Gamma_a$ from the cycling transition ($\Gamma_a=1/\tau_a$ is the decay rate of the transition) and reduce their intensity by about an order of magnitude to achieve efficient sub-Doppler cooling. At the end of the MOT loading cycle, we typically trap ~1.3 million atoms, within a radius of 0.26 mm. The atomic temperature is about $40$~$\mu$K, which is measured by monitoring the free expansion of the cloud using an electron-multiplying CCD camera. During the final 10 ms of the MOT loading cycle, we turn-off the hyperfine repumper beam. As a result, the atoms are optically pumped into the $F=2$ ground level at the end of the cycle. \\

{\it Dipole trap:} To form the dipole trap, we focus a far-off resonant laser beam at a wavelength of $1.055$~$\mu$m and overlap it with the MOT cloud. The dipole trapping beam comes from a separate laser system, which relies on a fiber-amplified ECDL. The optical power of the dipole trapping beam is 0.5 W, and it is focused to a spot size of $27$~$\mu$m ($1/e^2$ intensity radius), resulting in a trap-depth of $400$~$\mu$K. We typically transfer 11,000 atoms from the MOT to the dipole trap. The size of the atomic cloud confined to the dipole trap is $6.3$~$\mu$m~$\times~6.3$~$\mu$m~$\times~360$~$\mu$m (all dimensions are the $1/e$ density radius). 

After we load the dipole trap, we turn-off the MOT laser beams and keep the atoms trapped for 50 ms. At the end of 50 ms, we abruptly turn-off the dipole trapping beam, let the atomic cloud expand for a chosen amount of time, and then apply the 200 ns excitation pulse. The excitation laser is obtained by picking off a portion of the main MOT trapping laser, shifting its frequency using an AOM, and seeding a separate tapered amplifier. \\

{\it Michelson Interferometer}: The Michelson interferometer is formed by splitting the image of the fluorescing atomic cloud into two arms using a 50/50 beam-splitter. Each arm reflects off a mirror and the two arms are recombined back on the same beam-splitter. The recombined signal is collected using an aspheric lens coupled to a multi-mode fiber and detected with a photon counter. Each arm of the Michelson interferometer is about 5 cm long. Using this set-up, we essentially interfere two separate images of the ensemble, which can be transversely displaced from one another through misalignment of the mirrors. The interference of the two images can be characterized by precisely changing one of the arm lengths, which is achieved by using a piezo-electric transducer attached to one of the mirrors. The long-term phase stability of the interferometer between the two arms is quite good; the arm-lengths are typically stable to ~$\lambda/10$ over a 24-hour time window. 

The absolute phase of the emitted light from shot-to-shot is not determined and is essentially random. We note that the Michelson interferometer does not measure the absolute phase of the light. Rather, it is a measurement of the phase difference (relative phase) between the two arms of the interferometer (after the light is split into two with the beam-splitter). This relative phase is adjusted using a high-voltage piezo-electric transducer that moves one of the mirrors. 

In Fig. 6, we measure the contrast of the interference fringes as a function of the displacement between the two images. Here, we rotate one of the mirror knobs by a known amount, which then corresponds to certain transverse displacement of one of the arms right before the detector. The transverse displacement can be precisely calculated and is also verified by imaging a coherent laser beam with the interferometer. This transverse displacement is the horizontal axis of the two plots. Given a transverse displacement, we then move the piezo and record an interference pattern similar to that shown in Fig. 5. The fringe contrast is then calculated using this interference pattern. 

\section{Appendix B: Coherent versus incoherent emission}

In this section, we give a qualitative, but quite fundamental derivation of the figure-of-merit for the coherent emission rate from an ensemble, and contrast this figure-of-merit with that of incoherent emission. First consider an atomic cloud with $N$ atoms and a radial size of $R=\sqrt{\sigma_x^2 + \sigma_y^2 + \sigma_z^2}$. Here, the quantities $\sigma_x$, $\sigma_y$, and $\sigma_z$ are the radii of the cloud in the three spatial dimensions. The size of the cloud is much larger than the wavelength of the emitted light: $ R >> \lambda$. Now consider a specific atom in the ensemble, atom $i$, whose far field electric-field in the cloud would have the leading term, $E_i \sim 1/(kR) \sim  1/(R/\lambda)$ (the quantity $k=2 \pi /\lambda$ is the $k$-vector (wave-number) of the emitted radiation). 

For coherent emission, to find the total radiation rate, we would add up the electric fields due to each atom and square the result. In contrast, for incoherent emission, we would instead find the radiated intensity from each emitter (i.e., square their individual electric fields) and add up the intensities. In more concrete terms, for coherent emission, the decay quantities would scale as  $ \left( \sum_i^N E_i \right)^2 $, while for incoherent emission, they would scale as $  \sum_i^N E_i^2 $. Following this qualitative argument, we would expect the coherent emission rate to scale as $ \left( \sum_i^N E_i \right)^2  \sim  \left( \sum_i^N \frac{1}{kR} \right)^2 \sim \left( \frac{N}{R/\lambda} \right)^2$ For incoherent emission, the rate would instead scale as  $  \sum_i^N E_i^2 \sim \sum_i^N \frac{1}{(kR)^2} \sim \frac{N}{(kR)^2} \sim n \sigma R $ (i.e., the optical depth of the sample). All the incoherent processes, such as radiation trapping, would largely be determined by the optical depth. 

The observed subradiance time-scales in Fig. 4 qualitatively follows the figure-of-merit which is plotted in 4(b). As the density and the optical depth are increased, the observed subradiance increases and then plateaus, followed by a slight drop near the highest density. The portion of the data in the shaded region of Fig. 4(b) is especially significant. In this region, the atomic density varies by more than two orders of magnitude and the optical depth changes by more than an order of magnitude, and yet the observed subradiance is largely unchanged. This is a clear indication that the process is not dominated by the optical density or the optical depth, but instead by the figure-of-merit due to coherent emission.

The slight drop near the high-end of the data set is likely due to a currently unknown dephasing mechanism of the correlated states. The excited superradiant and subradiant states are essentially coherent superpositions of atoms across the ensemble, and the drop in the observed subradiance at such low densities may point to the fragile nature of these states. We note that dipole-dipole line-broadening coefficients for Rubidium are well-known. At the densities of our experiment, such broadening would be at least two orders of magnitude smaller than the radiative decay rate and would result in a negligible change of the linewidth. As a result, straightforward line broadening would not be able to explain the observed drop in subradiance. However, when correlated states across the ensemble contribute to the emission, even a small value of dipole-dipole broadening may be sufficient to dephase the established coherent emission. Our current hypothesis for the observed drop in subradiance is dephasing of the multi-atom correlations due to the density-dependent dipole-dipole interaction, similar to van der Waals dephasing that occurs at high densities, which is discussed in detail in Ref. \cite{haroche}. 

\section{Appendix C: Example decay curves} 

Figure~9 shows four of the decay curves that were used to obtain the data of Fig. 4. In each curve, the solid blue line is the experimental data and the red dashed line is a linear fit to the data to infer the $1/e$ decay time $\tau$, which is reported in Fig. 4. For comparison, the dashed black line is decay of an independent, isolated atom.

\begin{figure}[tbh]
\vspace{-1cm}
\begin{center}
\includegraphics[width=7cm]{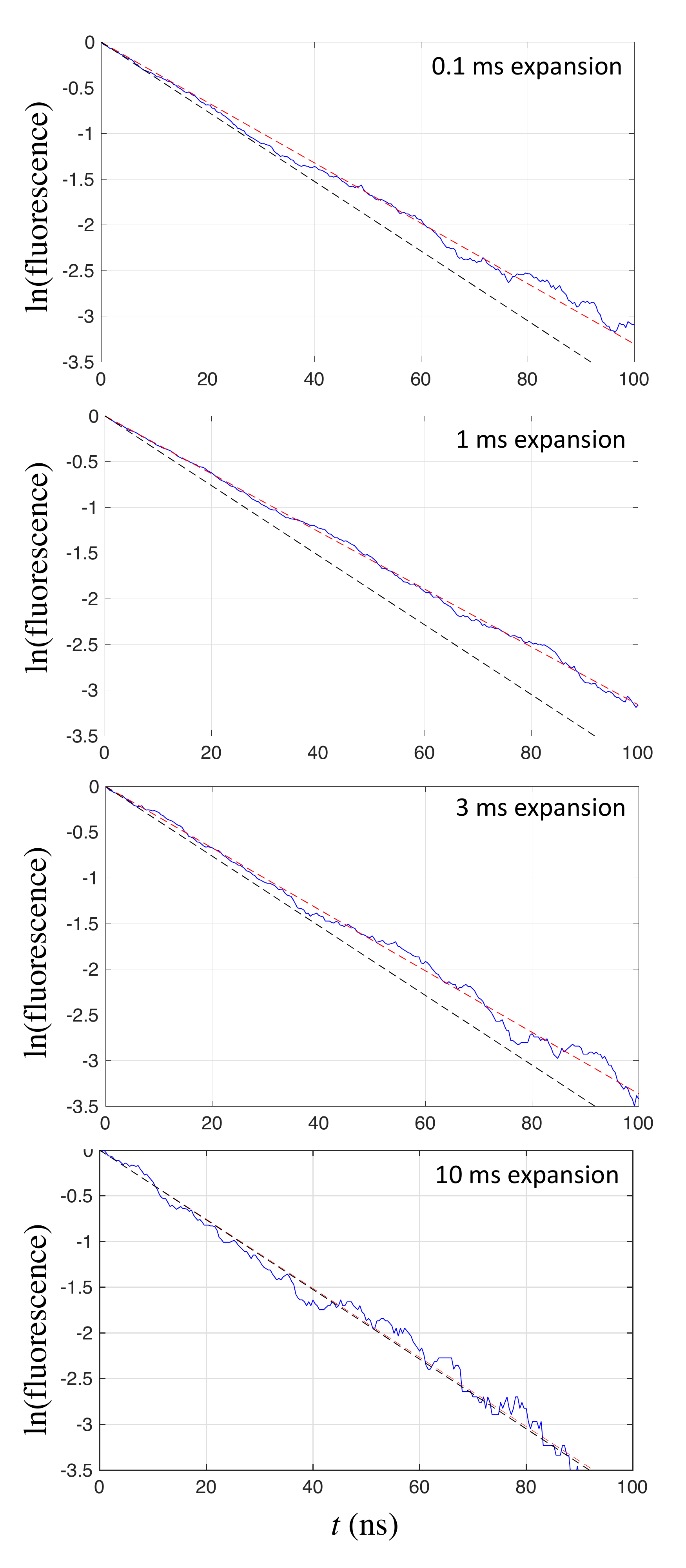}
\vspace{-0cm} 
\begin{singlespace}
\caption{\label{scheme} \small The decay of the ensemble (solid blue line) for cloud expansion times of 0.1, 1, 3, and 10~ms, respectively. In each plot. the dashed red line is a linear fit to the data, while the dashed black line is independent decay for reference. } 
\end{singlespace}
\end{center}
\vspace{-0.5cm}
\end{figure}

\section{Appendix D: Subradiance as a function of intensity of the excitation pulse}

All of the measurements reported in the main manuscript were performed in the strong excitation regime with a saturation parameter of $s/s_0 \approx 2$. However, we have also investigated the observed subradiance as a function of the saturation parameter. Figure~10 shows the measured $1/e$  decay time $\tau$ as the saturation parameter of the excitation laser is varied. While the early data points have large error bars due to reduced signal-to-noise, we observe that subradiance increases initially and then quickly reaches a plateau as the intensity of the excitation laser is increased. While we do not have a quantitative model to account for this behavior, qualitatively it is likely because, in this dilute regime, the initial excitation fraction (not just the number of atoms) plays a role. 

\begin{figure}[tbh]
\vspace{-0cm}
\begin{center}
\includegraphics[width=9cm]{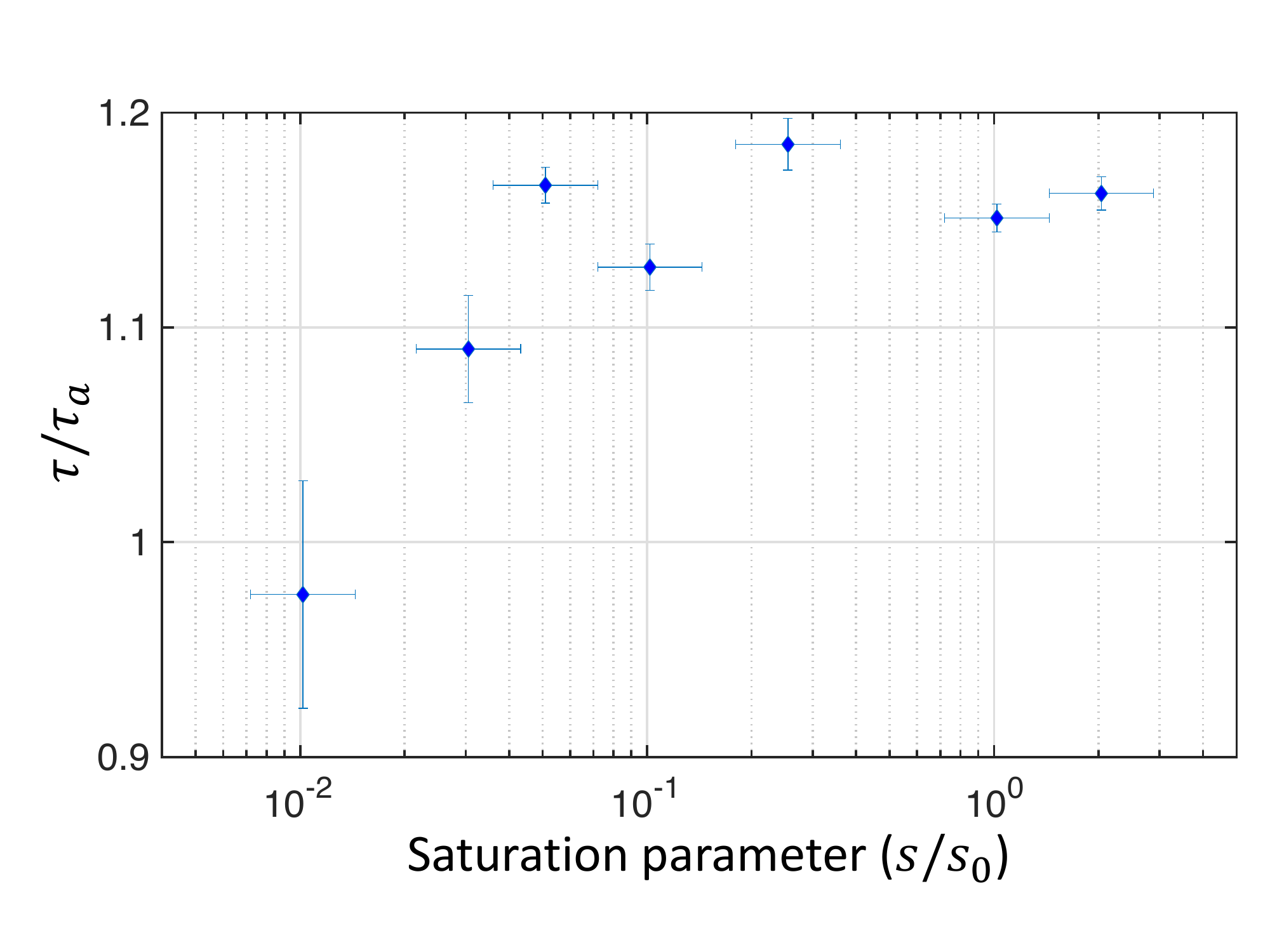}
\vspace{-0.5cm} 
\begin{singlespace}
\caption{\label{scheme} \small  The observed $1/e$ decay time $\tau$ as the saturation parameter of the excitation laser is varied. We observe that subradiance increases initially and then quickly reaches a plateau as the intensity of the excitation laser is increased. While we do not have a quantitative model to account for this behavior, qualitatively it is likely because, in this dilute regime, the initial excitation fraction (not just the number of atoms) plays a role. }
\end{singlespace}
\end{center}
\vspace{-0.3cm}
\end{figure}

We calibrate the saturation parameter for our excitation laser by probing the intensity of the total fluorescence as a function of the beam power. This calibration overcomes the experimental uncertainties such as the slight misalignment of the excitation laser beam from the atomic cloud. With the fluorescence measured, we then fit the data points using the well-known model for saturation of an atomic transition, $\rho_e = \frac{1}{2} \frac{s/s_0}{1+s/s_0}$. Here, the quantity $\rho_e$ is the excited state fraction. 

Figure~11 shows such a calibration for the excitation fraction of the atomic transition as a function of the optical power of the excitation laser. The solid red line is a fit to the data using the above saturation model. For all the experiments reported in the manuscript, the optical power of the excitation laser is 2 to 3 mW, corresponding to an excitation fraction of $\rho_e \approx 0.3$ (i.e., only 30\% of the population is in the excited level). We note, once again, that there is never population inversion in any of the data that is presented in the manuscript. 

\begin{figure}[tbh]
\vspace{-0cm}
\begin{center}
\includegraphics[width=10cm]{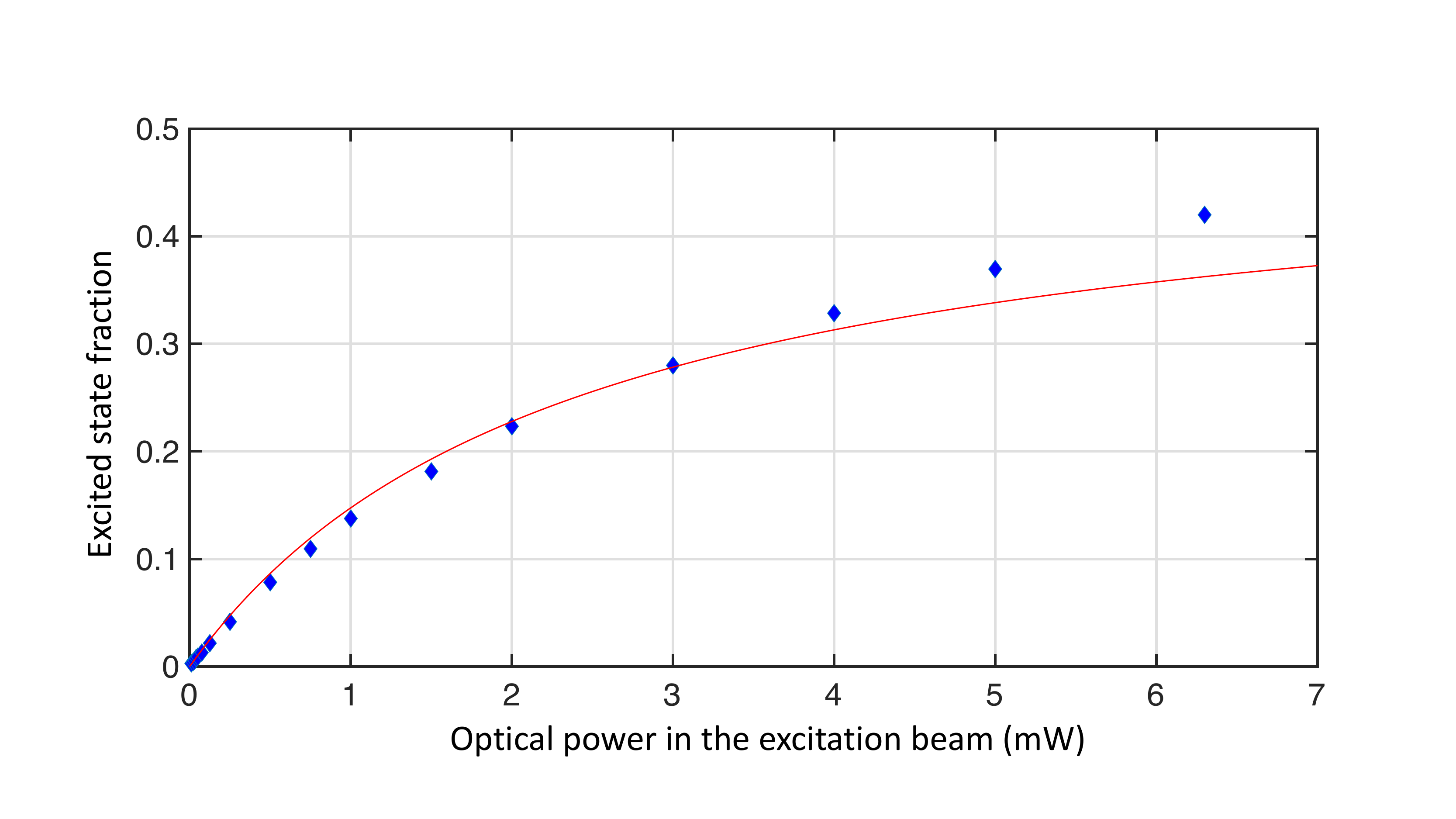}
\vspace{-0.5cm} 
\begin{singlespace}
\caption{\label{scheme} \small  The excitation fraction of the atomic transition as a function of the optical power in the excitation laser. The vertical error bars are negligible on this scale and are not shown. The solid red line is a fit to the data using the atomic saturation model (see text for more details).}
\end{singlespace}
\end{center}
\vspace{-0.3cm}
\end{figure}


\newpage


\begin{references}

\bibitem{dicke} R. H. Dicke, Coherence in Spontaneous Radiation Processes, Phys. Rev. {\bf 93}, 99 (1954).

\bibitem{haroche} M. Gross and S. Haroche, Superradiance: An Essay on the Theory of Collective Spontaneous Emission, Phys. Rep. {\bf 93}, 301 (1982).

\bibitem{scully1} M. O. Scully and A. A. Svdzinsky, The Super of Superradiance, Science {\bf 325}, 1510 (2009).

\bibitem{scully2}  M. O. Scully, E. S. Fry, C. H. Raymond Ooi, and K. Wódkiewicz, Directed Spontaneous Emission from an Extended Ensemble of N Atoms: Timing Is Everything, Phys. Rev. Lett. {\bf 96}, 010501 (2006).

\bibitem{feld}  N. Skribanowitz, I. P. Herman, J. C. MacGillivray, and M. S. Feld, Observation of Dicke Superradiance in Optically Pumped HF Gas, Phys. Rev. Lett. {\bf 30}, 309 (1973).

\bibitem{manassah} R. Friedberg, S. R. Hartmann, and J. T. Manassah, Frequency Shifts in Emission and Absorption by Resonant Systems of Two-level Atoms, Phys. Rep. {\bf 7}, 101 (1973).

\bibitem{kaiser1} W. Guerin, M. O. Araujo, and R. Kaiser, Subradiance in a Large Cloud of Cold Atoms, Phys. Rev. Lett. 116, 083601 (2016).

\bibitem{kaiser2} P. Weiss, M. O Araújo, R. Kaiser and W. Guerin, Subradiance and Radiation Trapping in Cold Atoms, New J. Phys. {\bf 20}, 063024 (2018).

\bibitem{browaeys} Giovanni Ferioli, Antoine Glicenstein, Loic Henriet, Igor Ferrier-Barbut , and Antoine Browaeys, Storage and Release of Subradiant Excitations in a Dense Atomic Cloud, Phys. Rev. X {\bf 11}, 021031 (2021). 

\bibitem{an} J. Kim, D. Yang, S. Oh, and K. An, Coherent Single-Atom Superradiance, Science 359, 662 (2018). 

\bibitem{gauthier} J. A. Greenberg and D. J. Gauthier, Steady-state, Cavityless, Multimode Superradiance in a Cold Vapor, Phys. Rev. A {\bf 86}, 013823 (2012). 

\bibitem{kuga} Y. Yoshikawa, Y. Torii, and T. Kuga, Superradiant Light Scattering from Thermal Atomic Vapors, Phys. Rev. Lett. {\bf 94}, 083602 (2005). 

\bibitem{thompson} J. G. Bohnet, Z. Chen, J. M. Weiner, D. Meiser, M. J. Holland, and J. K. Thompson, A Steady-State Superradiant Laser with Less-than One Intracavity Photon, Nature {\bf 484}, 78 (2012). 

\bibitem{molecules} B. McGuyer, M. McDonald, G. Iwata et al., Precise Study of Asymptotic Physics with Subradiant Ultracold Molecules, Nature Phys. {\bf 11}, 32 (2015).

\bibitem{ions} R. G. DeVoe and R. G. Brewer, Observation of Superradiant and Subradiant Spontaneous Emission of Two Trapped Ions, Phys. Rev. Lett. {\bf 76}, 2049 (1996).

\bibitem{atoms} D. Pavolini, A. Crubellier, P. Pillet, L. Cabaret, and S. Liberman, Experimental Evidence for Subradiance, Phys. Rev. Lett. {\bf 54}, 1917 (1985).

\bibitem{nanofiber} P. Solano, P. Barberis-Blostein, F. K. Fatemi et al., Super-radiance Reveals Infinite-range Dipole Interactions Through a Nanofiber, Nat. Commun. {\bf 8}, 1857 (2017).

\bibitem{metamaterial} S. D. Jenkins, J. Ruostekoski, N. Papasimakis, S. Savo, and N. I. Zheludev, Many-Body Subradiant Excitations in Metamaterial Arrays: Experiment and Theory, Phys. Rev. Lett. {\bf 119}, 053901 (2017).

\bibitem{bec} P. Wolf, S.C. Schuster, D. Schmidt, S. Slama, and C. Zimmermann, Observation of Subradiant Atomic Momentum States with Bose-Einstein Condensates in a Recoil Resolving Optical Ring Resonator, Phys. Rev. Lett. {\bf 121}, 173602 (2018).

\bibitem{havey} S.J. Roof et al., Observation of Single-Photon Superradiance and the Cooperative Lamb Shift in an Extended Sample of Cold Atoms, Phys. Rev. Lett. {\bf 117}, 073003 (2016).

\bibitem{diamond1} C. Bradac, M. T. Johnsson, M. V. Breugel, et al., Room-temperature Spontaneous Superradiance From Single Diamond Nanocrystals, Nat. Commun. {\bf 8}, 1205 (2017).

\bibitem{diamond2} A. Angerer, K. Streltsov, T. Astner, T. et al., Superradiant Emission From Color Centers in Diamond, Nat. Phys. {\bf 14}, 1168 (2018).

\bibitem{superconducting} Z. Wang et al., Controllable Switching Between Superradiant and Subradiant States in a 10-qubit Superconducting Circuit, Phys. Rev. Lett. {\bf 124}, 013601 (2020).

\bibitem{eberly} J. H. Eberly, Emission of One Photon in an Electric Dipole Transition of One Among N Atoms, J. Phys. B {\bf 39}, S599 (2006).

\bibitem{scully3} M. O. Scully, Single Photon Subradiance: Quantum Control of Spontaneous Emission and Ultrafast Readout, Phys. Rev. Lett. {\bf 115}, 243602 (2015).

\bibitem{adams} R. J. Bettles, S. A. Gardiner, and C. S. Adams, Cooperative Eigenmodes and Scattering in One-Dimensional Atomic Arrays, Phys. Rev. A {\bf 94}, 043844 (2016). 

\bibitem{zoller} P.-O. Guimond, A. Grankin, D.V. Vasilyev, B. Vermersch, and P. Zoller, Subradiant Bell States in Distant Atomic Arrays, Phys. Rev. Lett. {\bf 122}, 093601 (2019).

\bibitem{jenkins} S. D. Jenkins and J. Ruostekoski, Controlled Manipulation of Light by Cooperative Response of Atoms in an Optical Lattice, Phys. Rev. A {\bf 86}, 031602 (2012).

\bibitem{ritsch} H. Zoubi and H. Ritsch, Lifetime and Emission Characteristics of Collective Electronic Excitations in Two-Dimensional Optical Lattices, Phys. Rev. A {\bf 83}, 063831 (2011).

\bibitem{zanthier} D. Bhatti, R. Schneider, S. Oppel and J. von Zanthier , Directional Dicke Subradiance with Nonclassical and Classical Light Sources, Phys. Rev. Lett. {\bf 120}, 1136 (2018).

\bibitem{agarwal} R. Wiegner, J. von Zanthier, and G. S. Agarwal, Quantum-interference-initiated Superradiant and Subradiant Emission from Entangled Atoms, Phys. Rev. A {\bf 84}, 023805 (2011).

\bibitem{kimble} A. Asenjo-Garcia, M. Moreno-Cardoner, A. Albrecht, H.J. Kimble, and D. E. Chang, Exponential Improvement in Photon Storage Fidelities Using Subradiance and “Selective Radiance” in Atomic Arrays, Phys. Rev. X {\bf 7}, 031024 (2017).

\bibitem{yelin} E. Shahmoon, D. S. Wild, M. D. Lukin, and S. F. Yelin, Cooperative Resonances in Light Scattering from Two-Dimensional Atomic Arrays, Phys. Rev. Lett. {\bf 118}, 113601 (2017).

\bibitem{bloch} J. Rui, D. Wei, A. Rubio-Abadal, S. Hollerith, J. Zeiher, Dan M. Stamper-Kurn, C. Gross, and I. Bloch, A Subradiant Optical Mirror Formed by a Single Structured Atomic Layer, Nature {\bf 583}, 369 (2020).

\bibitem{dipto} D. Das, B. Lemberger, and D. D. Yavuz, Subradiance and Superradiance-to-Subradiance Transition in Dilute Atomic Clouds, Phys. Rev. A {\bf 102}, 043708 (2020).

\bibitem{induction} M. Afzelius et al., Interference of Spontaneous Emission of Light From Two Solid-State Ensembles, New J. of Phys. {\bf 9}, 413 (2007). 

\bibitem{ben} B. Lemberger and D. D. Yavuz, Effect of Correlated Decay on Fault-tolerant Quantum Computation, Phys. Rev. A {\bf 96}, 062337 (2017).

\bibitem{coherence} Z. Bialynicka-Birula, Coherence of the Radiation from the Superradiant State, Phys. Rev. D {\bf 1}, 400 (1970).  

\bibitem{kimble_review} D. E. Chang, J. S.  Douglas, A. Gonzalez-Tudela, A., C. L. Hung, and H. J. Kimble, {\it Colloquium: Quantum Matter Built from Nanoscopic Lattices of Atoms and Photons},  Rev.
Mod. Phys. {\bf 90}, 031002 (2018).

\bibitem{NielsenChuang} M. A. Nielsen and I. L. Chuang, {\it Quantum Computation and Quantum Information} (Cambridge University Press, 2000).

\bibitem{neutral_review1} M. Saffman, T. G. Walker, and K. Molmer, {\it Quantum Information with Rydberg Atoms}, Rev. Mod. Phys. {\bf 82}, 2313 (2010).


\bibitem{burkard} B. M. Terhal and G. Burkard, Fault-tolerant Quantum Computation for Local Non-Markovian Noise, Phys. Rev. A {\bf 71}, 012336 (2005).

\bibitem{preskill} D. Aharonov, A. Kitaev, and J. Preskill, Fault-Tolerant Quantum Computation with Long-Range Correlated Noise, Phys. Rev. Lett. {\bf 96}, 050504 (2006). 

\bibitem{mucciolo1} E. Novais and E. R. Mucciolo, Surface Code Threshold in the Presence of Correlated Errors, Phys. Rev. Lett. {\bf 110}, 010502 (2013). 

\bibitem{mucciolo2} P. Jouzdani, E. Novais, I. S. Tuptsyn, and E. R. Mucciolo, Fidelity Threshold of the Surface Code Beyond Single-Qubit Error Models, Phys. Rev. A {\bf 90}, 042315 (2014). 



\end{references}
\end{document}